\documentclass[aps,twocolumn,
prb,
showpacs]{revtex4}

\usepackage{dsfont}
\usepackage{graphpap}
\usepackage[dvips]{graphicx}
\usepackage[dvips]{graphics}
\usepackage{amsmath}
\usepackage{amsbsy}
\usepackage{amssymb}
\usepackage{epic}
\usepackage{epsfig}    
\usepackage{color}

\newcommand{\expval}[1]{\left< #1 \right>}
\newcommand{\ket}[1]{\left| #1 \right>}
\newcommand{\bra}[1]{\left< #1 \right|}

\newcommand{\lB}{l_B}

\newcommand{\pp}{n_i}

\newcommand{\KKp}{$K\!-\!K^{\prime}$}

\begin{document}
\title{Transition to Landau Levels in Graphene Quantum Dots}
 \author{F. Libisch$^{1}$,\footnote{Corresponding author, e-mail:
                              florian@concord.itp.tuwien.ac.at} ,
   S. Rotter$^{1}$, J. G\"uttinger$^{2}$, C. Stampfer$^{2,3}$, and
   J. Burgd\"orfer$^1$} \affiliation{$^1$Institute for Theoretical
   Physics, Vienna University of Technology\\Wiedner Hauptstra\ss e
   8-10/136, A-1040 Vienna, Austria, EU \\ $^2$Solid State
   Physics Laboratory, ETH Zurich, 8093 Zurich, Switzerland \\ $^3$
 JARA-FIT and II. Institute of Physics, RWTH Aachen, 52074 Aachen,
 Germany, EU}\date{ \today}

\begin{abstract}
We investigate the electronic eigenstates of graphene quantum dots of
realistic size (up to 80~nm diameter) in the presence of a
perpendicular magnetic field $B$. Numerical tight-binding calculations
and Coulomb-blockade measurements performed near the Dirac point
exhibit the transition from the linear density of states at $B=0$ to
the Landau level regime at high fields. Details of this transition
sensitively depend on the underlying graphene lattice structure, bulk
defects, and localization effects at the edges. Key to the
understanding of the parametric evolution of the levels is the
strength of the valley-symmetry breaking \KKp\ scattering. We show
that the parametric variation of the level variance provides a
quantitative measure for this scattering mechanism. We perform measurements of
the parametric motion of Coulomb blockade peaks as a function of
magnetic field and find good agreement. We demonstrate that the
magnetic-field dependence of graphene energy levels may serve as
a sensitive indicator for the properties of graphene quantum dots and,
in further consequence, for the validity of the Dirac-picture.

\end{abstract}


\maketitle 

\section{Introduction}

Graphene nanostructures\cite{ han07, che07, dai08, sta08a, pon08,
  gue08, sch09, tod09, liu09, rit09,mos09,sta09} attract increasing
attention mainly due to their potential applications in high mobility
electronics\cite{gei07,kat07} and solid state quantum information
processing.\cite{tra07} In particular, low nuclear spin concentrations
expected in graphene promise long spin
lifetimes\cite{kan05,min06,hue06,tra07} and make graphene quantum dots
(QDs)\cite{sta08a,pon08,gue08,sch09} interesting for spin-qubit
operations.\cite{tra07} Moreover, graphene nanostructures may allow to
investigate phenomena related to massless Dirac Fermions in reduced
dimensions.\cite{pon08,ber87,sch08,rec07, rec09,Wimmer,lib09,you09} Intensive
research has been triggered by the unique electronic properties of
graphene~\cite{cas08} including the gapless linear dispersion, and the
Landau level (LL)
spectrum.\cite{Haldane88,Zheng02,McCann06,Sheng07,Ostrovsky08,nov05,zha05,Schulz09a, Schulz09b}
Recent advances in fabricating width-modulated graphene nanoribbons
have helped to overcome intrinsic difficulties in creating tunneling
barriers and confining electrons in graphene, where transport is
dominated by Klein tunneling-related phenomena.\cite{dom99,kat06}
Graphene QDs have been fabricated and Coulomb
blockade,\cite{sta08a,pon08} quantum confinement\cite{sch09} and
charge detection\cite{gue08} have been demonstrated.

\begin{figure}
\hbox{}\hfill\epsfig{file=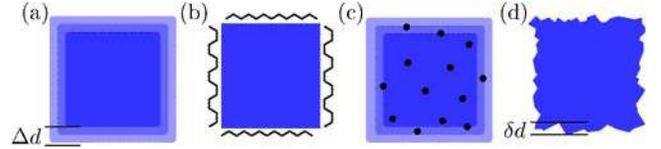, width = 8.5cm}\hfill\hbox{}
\caption{Shapes and sizes (50$\times$50~nm) of graphene quantum dots
  confined by (a) a smooth valleyspin-conserving
  potential [Eq.~(\ref{eq:pot}), the length scale of the confinement
    is marked by $\Delta d$], (b) atomically sharp armchair and zigzag
  boundaries. Dots with disorder due to (c) bulk defects or (d) edge
  roughness.}
\label{fig:geom}
\end{figure}

In this article, we focus on the eigenenergies of graphene quantum
dots (see Fig.~\ref{fig:geom}) as a function of a perpendicular
magnetic field. In graphene, the linear band crossing at the so-called
Dirac point suggests a close connection between the dynamics of electrons
and free, ultrarelativistic Dirac particles.\cite{Semenoff} One
might therefore expect a magnetic-field dependence of quantum dot
eigenenergies that closely mirrors that of massless Dirac particles.
Indeed, this connection has been used recently to discuss the spectrum
of ideal, circular graphene dots with smooth confinement.\cite{sch08,
  rec09} However, in more realistic models of finite graphene
nanostructures, quantum confinement, edge effects, and lattice defects
introduce a host of competing length scales absent from the simple
Dirac picture. Much progress has been made in understanding the unique
LL spectrum, and the resulting Hall effect, in
graphene.\cite{Haldane88,Zheng02,McCann06,Sheng07,Ostrovsky08,Schulz09a, Schulz09b} The
magnetic-field dependence of the addition spectrum has been exploited
in recent work to (approximately) pin down the electron-hole crossover
point.\cite{Guttinger09} In the present paper we report on a
systematic study of the $B$-field dependence of electronic eigenstates
of graphene quantum dots of experimentally realizable size (diameter
$d\le 80$~nm). We highlight the interplay of different length scales
controlling the break-down of the valley symmetry by
\KKp\ scattering. The latter is found to be key to the understanding
of the diamagnetic spectrum. We find the $B$-field dependence of the
level variance to be a sensitive measure for the strength of
\KKp\ scattering and obtain good agreement with experimental Coulomb
blockade data.

The paper is organized as follows: we first briefly summarize the
Dirac picture of Landau level formation for massless charged Dirac
particles and discuss the length scales relevant to its applicability
to finite-size graphene quantum dots (Sec.~II). In Sec.~III we present
realistic simulations for graphene quantum dots with zigzag and
armchair edges, with edge roughness as well as with bulk disorder. A comparison
between the calculated $B$-field dependence of the level variance and
experimental data is given in Sec.~IV, followed by a short summary
(Sec.~V).

\section{The Dirac picture and its limitations}

The remarkable similarity of the low-energy band structure of graphene
with the dispersion relation of a massless Dirac particle in two
dimensions has been widely exploited in a variety of theoretical
models for graphene.\cite{cas08} However, the applicability of such
models requires careful consideration of competing effects that go
beyond the simple, yet intriguing Dirac picture.\cite{McCann06b,
  Sasaki06} A case in point is the diamagnetism, i.e., the
magnetic response of a finite-size graphene quantum dot. It is of
considerable interest to inquire into the applicability as well as the
limitations of the well-known diamagnetic theory of charged massless
Dirac fermions.

The magnetic-field ($B$) dependence of the spectrum of free Dirac
particles was first solved in an early paper by Rabi\cite{Rabi}
shortly after the Dirac equation was proposed. The Dirac equation for
a massless particle with charge $q (=-\left|e\right|)$ in the presence
of a potential $V(\mathbf{x})$ with time-like coupling and a perpendicular,
homogeneous magnetic field $\mathbf{B} = \nabla \times \mathbf{A} =
-By\nabla\times\mathbf{e_x}$ reads
\begin{equation}
H_D = H_0 + H_B = v_{\mathrm{F}}\vec{\sigma}\cdot\left(\vec p - \frac{q}{c}\vec A \right) + 
\sigma_0 V(\vec x)\label{eq:Hdirac},
\end{equation}
with $\vec{\sigma}=(\sigma_x,\sigma_y)$ the Pauli matrices and
$\sigma_0 = \mathds{1}$.
In the limiting case of strong magnetic field where $|qA/c| \gg
|V(\mathbf x)|$ the solution of Eq.~(\ref{eq:Hdirac}) predicts the
formation of Landau levels,\cite{McClure56, cas08, rec07, rec09, sch08}
\begin{equation}
E_n^D (B)= \mathrm{sgn}(n)\sqrt{2\left|e\right|\hbar v_{\mathrm{F}}^2
  |n| B},\quad n \in \mathbb{Z}_0.
\label{B_D_L}
\end{equation}
We explicitly label this reference spectrum with the superscript ``D''
(for Dirac equation).
Equation (\ref{B_D_L}) contains several remarkable features absent
from non-relativistic diamagnetism: a ground state Landau level $n=0$
the energy of which does not depend on $B$ at all. Higher Landau
levels $n=\pm 1, \pm 2,\ldots$, are distributed symmetrically around
$n=0$, and feature a $\sqrt B$ rather than a linear dependence on $B$
known from non-relativistic diamagnetism. The high-field regime
[Eq.~(\ref{B_D_L})] is controlled by just two length scales, the
(energy dependent) de Broglie wavelength $\lambda_{\mathrm{F}}$ and
the magnetic length $l_B = \sqrt{\hbar c/(eB)}$.  The strong (weak)
field regime is characterized by $l_B \ll \lambda_{\mathrm F}$ ($l_B
\gg \lambda_{\mathrm F}$). In the limit of weak magnetic fields,
Eq.~(\ref{eq:Hdirac}) predicts the lowest-order energy corrections to
scale linearly with $B$, unlike the conventional non-relativistic
behavior ($\propto B^2$). Eigenstates of Eq.~(\ref{eq:Hdirac})
form two-spinors with definite helicity (or ``chirality'')
\begin{equation}
\hat h\ket\psi = \frac 12 \vec\sigma\cdot\frac{\vec p}{|p|}\ket{\psi} = \pm \frac 12 \ket\psi
\end{equation}
in the absence of external fields.

The ideal, infinitely extended graphene sheet featuring a honeycomb
lattice made up by two interleafed triangular sublattices (A and B),
can be described in nearest neighbor tight-binding approximation by
the Hamiltonian\cite{Wallace47}
\begin{equation}
  H = \sum_{i,s}\ket{\phi_{i,s}}V_i\bra{\phi_{i,s}}-t \sum_{(i,j),s}\ket{\phi_{i,s}}\bra{\phi_{j,s}} + h.c.\,, \label{H_Graph_TB}
\end{equation}
where the sum $(i,j)$ extends over pairs of adjacent lattice sites,
 $\ket{\phi_{j,s}}$ is the tight-binding orbital 
at lattice site $j$, $V_i$ is a locally varying potential, and $t$ (of
the order of $2.8$~eV) is the nearest neighbor hopping matrix element. [In the
  numerical calculations we take into account second and third nearest
  neighbor coupling \cite{TBReich, TBWirtz, lib09} in addition to
  Eq.~(\ref{H_Graph_TB}) in order to quantitatively account for the
  realistic band structure.] Close to the Fermi energy, the band
structure of Eq.~(\ref{H_Graph_TB}) can be approximated (assuming that $V_i \ll t$)
by a conical dispersion relation around the $K$
point,\cite{Semenoff}
\begin{equation}\label{Eq:Semenoff}
  E(k + k_K) = E(k_K) + k \partial_k E(k_K) +
  \mathcal{O}(k_K^2)\approx v_{\mathrm{F}}|k|,
\end{equation}
where we have set $E(k_K) = 0$. Note that the above expansion ignores
both the length scale of the graphene lattice constant $a=1.4$
\AA\ and preferred directions of the lattice: due to the discrete
lattice symmetry, the cone structure becomes squeezed along the
\KKp\ directions, an effect known as triangular
warping.\cite{McCann06b, cas08} More importantly, the band structure
features \emph{two} non-equivalent cones (``valleys'') at the $K$ and
$K'$ points in the reciprocal lattice. This additional degeneracy
allows to formaly represent the low-energy band-structure near $E=0$
in terms of Dirac-like four-spinors
$\ket\psi=(\psi_A^{(K)},\psi_B^{(K)}, \psi_A^{(K')},\psi_B^{(K')})$
with amplitudes for the A-B sublattice in real space and for the K-K' valleys in
reciprocal space. Operators in the four-spinor space can be represented
by tensor products of $(\sigma_0,\vec \sigma)$ matrices acting on A-B
sublattice amplitudes and analogous $(\tau_0,\vec\tau)$ Pauli-matrices
acting on \KKp\ amplitudes. Choosing the origin in $k$-space such that
the connecting line between $K$ and $K'$ is along $y$, the effective
Dirac Hamiltonian in the absence of external scalar potentials becomes\cite{Slonc}
\begin{equation}
H_0 =  
\vec\sigma\cdot(\vec p - \frac{q}{c}\vec A)\otimes \tau_1 +
\vec\sigma^*\cdot(\vec p - \frac{q}{c}\vec A)\otimes \tau_2,\label{E_Graph}
\end{equation}
where $\tau_{1,2} = (\tau_0 \pm \tau_z)/2$. 
In addition to chirality, the valley-pseudospin projection
\begin{equation}
\tau_z\ket{j}=j\ket{j},\quad j = \pm\frac 12,
\end{equation}
associated with the valley degree of freedom is conserved. The upper
(``particle-like'', $E > 0$) and lower (``hole-like'', $E < 0$) cones
touching each other at $K$ and $K'$ with $E=0$ are related to each
other by a particle-hole transformation
\begin{equation}
\hat C = \sigma_z \otimes \tau_0, \quad \hat C H \hat C^{-1} = -H.\label{eq:partholesymm}
\end{equation}
In the presence of a time-like scalar potential $V(\vec x) \sigma_0
\otimes \tau_0$, the Hamiltonian is invariant under an anti-unitary
transformation (``time reversal''), $\hat T = i \sigma_y \otimes
\tau_0 \;\mathcal C$, where $\mathcal C$ denotes complex
conjugation.\cite{cas08} The wavefunctions at $K$ and $K'$ are related
by time reversal symmetry. This symplectic symmetry ($T^2=-1$) is
broken in the presence of a magnetic field
\begin{equation}
\hat T H(\vec A) \hat T^{-1} = H(-\vec A)
\end{equation}
lifting the two-fold Kramers-like degeneracy (Note that physical spin is
not included in the present analysis.). 

\begin{figure}
\hbox{}\hfill\epsfig{file=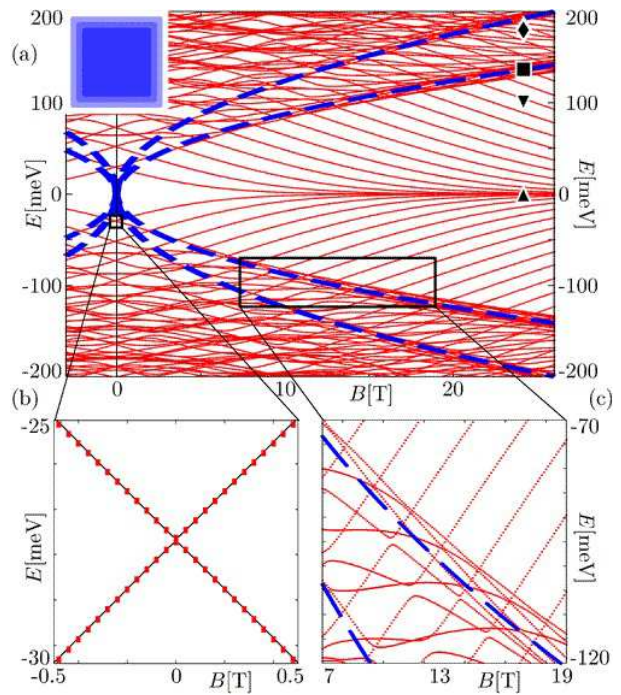, width = 8cm}\hfill\hbox{}
\caption{(color online) (a) Magnetic-field dependence of the
  eigenenergies of a graphene quantum dot with smooth confinement
  which approximately preserves valley symmetry
  [Eq.~(\ref{eq:pot}), see Fig.~\ref{fig:geom}(a)]. Landau levels for
  $n=\pm 1,\pm 2$ [dashed lines, see Eq.~\eqref{B_D_L}] are inserted as
  guide to the eye. The four symbols
  ($\blacktriangle,\blacktriangledown,\blacksquare,\blacklozenge$)
  mark parameter values for which eigenstates are shown in
  Fig.~\ref{fig:ES_clean}. (b) Close-up of the avoided crossing of two
  eigenstates in (a).  Dots represent numerical data, the continuous
  line is a fit to Eq.~(\ref{E_pert}). (c) Close-up of avoided
  crossings around the diabatic ridge formed by the first Landau level
  (see text).}
\label{fig:EV_clean}
\end{figure}

We now consider a finite-size system of linear dimension $d$, where
$V(\mathbf x)$ takes on the role of a confinement potential. With this
additional length scale present, the Landau-level solution
[Eq.~(\ref{B_D_L})] is only valid in the strong magnetic field regime
with $\lB \ll d$, while in the weak field regime, $\lB \gg d$, the
spectrum will be determined by $V(\mathbf x)$. For zero magnetic
field, eigenstates $\ket{\psi_K}$ and $\ket{\psi_{K'}}$ localized at
the $K$ and $K'$ points are degenerate. Turning on a magnetic field
lifts this degeneracy without (to lowest order) introducing couplings
between $K$ and $K'$. Following first-order degenerate
Rayleigh-Schr\"odinger perturbation theory, the perturbation matrix
$W$ describing the lowest order correction to the field-free spectrum
of Eq.~(\ref{E_Graph}) takes the form [for the gauge $A = (-By,0,0)$]
\begin{equation}
W = \expval{H_B} \sigma_x\otimes \tau_z,\label{E_pert}
\end{equation}
where $\expval{H_B}_{KK} := Bq \bra{\psi_B} y \ket{\psi_A} / c$ is
real, and linear in $B$. The perturbation preserves the valley
symmetry. The absence of \KKp\ coupling and the linear magnetic field
dependence of $\expval{H_B}_{KK}$ for each decoupled Dirac cone
implies that energy eigenvalues \emph{linearly} cross the line $B=0$
in pairs of two, forming an $x$-shaped intersection (see
Fig.~\ref{fig:EV_clean}). This is in contrast to the nonrelativistic
diamagnetic response $\propto B^2$ in the perturbative limit. It
rather resembles the paramagnetic level splitting in conventional
quantum dots when the magnetic field lifts a degeneracy.  Examples of
the latter are lifting of Kramers' degeneracy \cite{adam02} or a
symmetry--induced degeneracy as in circular quantum dots.\cite{lent91}

In order to quantitatively simulate the diamagnetic response of a
finite-size graphene quantum dot we first consider a smooth
confinement potential that is slowly varying on a length scale of the
lattice constant such as to approximately conserve the valley-pseudospin
projection $\tau_z$ (or valley symmetry). Moreover, it conserves
particle-hole (anti-)symmetry [Eq.~(\ref{eq:partholesymm})] in order to prevent Klein
tunneling. Such a potential was first proposed by Berry and 
Mondragon\cite{ber87} in the context of neutrino billiards,
\begin{equation}\label{eq:pot}
V(\mathbf x) = V_0 (e^{\Delta r(\mathbf x)/\Delta d}-1)
\sigma_z\,,
\end{equation}
where $\Delta r(\mathbf x)$ is the outward distance from the quantum
dot boundary and $\Delta d$ introduces an additional length scale
controlling the preservation of valley symmetry. We choose $\Delta
d=24$\AA\ [see Fig.~\ref{fig:geom}(a)] much larger than the lattice
spacing ($\Delta d \gg a\approx 1.42$\AA). Consequently,
Eq.~(\ref{eq:pot}) varies slowly on the scale of the lattice constant,
conserves valley symmetry [to order $(a/\Delta d)^2$] and provides a
realization of the (approximately) \KKp\ decoupled diamagnetic
perturbation [Eq.~(\ref{E_pert})]. We note that realizations of
potentials of the form of Eq.~(\ref{eq:pot}) are, to our knowledge,
currently experimentally not available. We employ a third-nearest
neighbor\cite{Sasaki06, TBReich, TBWirtz} tight-binding approximation (to
correctly describe triangular warping) and simulate a
50$\times$50~nm graphene QD containing $\approx$~100.000 carbon
atoms. The magnetic field is included by a Peierls phase factor. We
use a Lanczos diagonalization in conjunction with an LU factorization
to efficiently calculate the 500 eigenvalues closest to the Fermi
edge\cite{Lanczos, MUMPS} [see Fig.~\ref{fig:EV_clean}].

In the limit of weak magnetic fields, we find that our numerical
results, indeed, follow the linear $B$-field dependence of the energy
eigenvalues as predicted by perturbation theory [Eq.~(\ref{E_pert})]
[see Fig.~\ref{fig:EV_clean}(b)]. Residual deviations from the perfect
lattice symmetry (due to the finite width $\Delta d$ of the
confinement) and, thus, weak non-conservation of $\tau_z$ appear as
minute energy splittings between near-degenerate levels in the $B\to0$
limit. The resulting level splitting at $B=0$ is, however, very small
($120\mu$~eV), i.e., two orders of magnitude below the mean level
spacing ($\approx 10$~meV).

Turning now to the high-field limit, $\lB \ll d$, the influence
of confinement effects should be diminished and the formation of
Landau levels following the Dirac picture [Eq.~(\ref{eq:Hdirac})] is
expected. The transition from low to high magnetic fields drastically
changes the density of states (DOS). The depletion of the DOS near
$E=0$ at low fields,
\begin{equation}
\rho(E) = \frac{d^2}{2(\hbar v_{\mathrm F})^2} \left|E\right|,
\end{equation}
is replaced, for increasing $B$, by an increasing number of eigenstates
moving towards the Landau level at $E^D_0$, which is located at
the Dirac point [see Fig.~\ref{fig:EV_clean}(a)]. More specifically,
all graphene levels with energies in between the two first Landau
levels, $E^D_{-1}<E<E^D_1$, adiabatically converge to the level at
$E^D_0=0$. As we have shown
recently,\cite{Guttinger09} this unique feature can be used to pin
down the energetic position of the Dirac point in the experiment and thus of
the electron-hole crossover region in real graphene quantum dot devices.

\begin{figure}
\hbox{}\hfill\epsfig{file=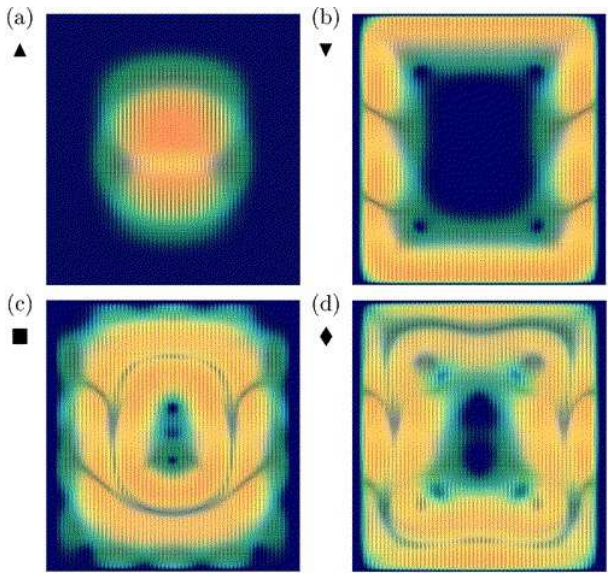, width = 8cm}\hfill\hbox{}
\caption{(color online) Typical eigenstates (plotted is the absolute
  square of the wavefunction) of a graphene quantum dot with smooth
  confinement [see Fig.~\ref{fig:geom} (a) and Eq.~(\ref{eq:pot})] at
  high magnetic field (B=25 T), corresponding to the zeroth [(a) and
    (b)] and first [(c) and (d)] Landau level. Symbols
  ($\blacktriangle,\blacktriangledown,\blacksquare,\blacklozenge$)
  correspond to those marking the position of these states in the
  energy level diagram [Fig.~\ref{fig:EV_clean}(a)].}
\label{fig:ES_clean}
\end{figure}

While, at low fields, the valley symmetry is approximately preserved by the
potential in Eq.~(\ref{eq:pot}), a large number of
sizeable avoided crossings appear at higher magnetic fields as the
magnetic length is reduced to $\lB \ll d$. Edge states that
couple to bulk states or to other edge states become prevalent.  The
complicated pattern in Fig.~\ref{fig:EV_clean}(a),(c) of many avoided
crossings near the first Landau energy $E=E^D_{\pm 1}$ reflects this
interplay between magnetic bulk and edge states.  Levels with
eigenenergies that follow the predicted values for the Landau levels,
$E^D_n$, are localized in the interior of the quantum dot and well
separated from the edges. Conversely,
the states with energies in between the values $E^D_n$ should be
strongly influenced by the spatial confinement in the quantum
dot. Wave functions corresponding to energy levels close to $n=0$ and
$n=1$ Landau levels [Fig.~\ref{fig:ES_clean}(a,c)] as well as those
in between $n=0$ and $n=\pm1$ [Fig.~\ref{fig:ES_clean}(b)] and between
$n=1$ and $n=2$ [Fig.~\ref{fig:ES_clean}(d)] confirm these
expectations.

Apparently, the typical level splittings at the avoided crossings
[Fig.~\ref{fig:EV_clean}(b,c)] are dramatically enhanced in the
high-field regime. This is due to the fact that, as compared to the
low-field case, the amplitudes of wave functions are enhanced at the
dot boundary [see Fig.~\ref{fig:ES_clean}(b)]. Since, in addition, the
lattice symmetry is broken at the boundary, these edge states do
have an increased coupling strength to all other states in the
sample. Following the Wigner-von Neumann non-crossing
rule\cite{wignervonneumann} this increased coupling strength leads to
increased level splittings at the crossing point.

\section{Realistic graphene quantum dots}

\subsection{Clean dots with sharp edges}

\begin{figure}
\hbox{}\hfill\epsfig{file=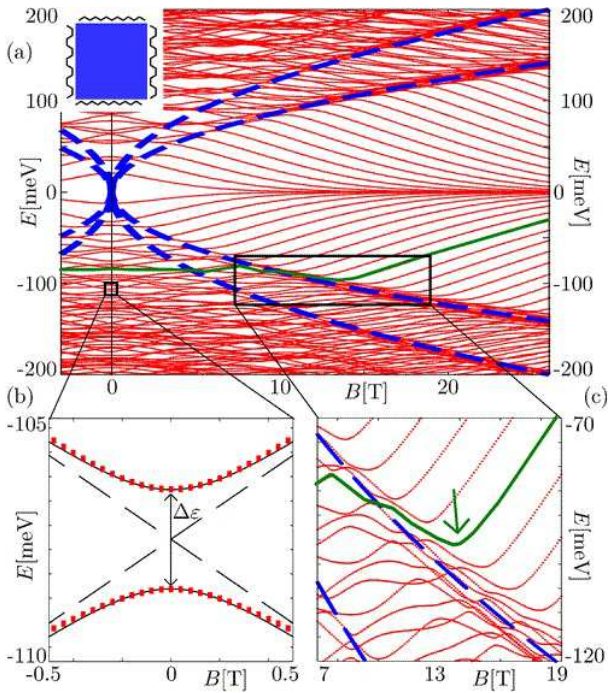, width = 8cm}\hfill\hbox{}
\caption{(color online) Same as Fig.~\ref{fig:EV_clean}, but for a
  quantum dot with atomically sharp zigzag and armchair
  boundaries. The solid line in (b) is a fit to Eq.~(\ref{EV_pert_kk}), the
  dashed line (corresponding to $V_{KK'} = 0$) is inserted as guide to the eye. The
  evolution of one eigenenergy with magnetic field is drawn with a
  thick green line as guide to the eye in (a), (c). The arrow in (c) marks a
  kink in the magnetic field dependence of this state (see text).}
\label{fig:EV_hard}
\end{figure}

We turn now to realistic graphene quantum dots where the
nanostructures are terminated by atomically sharp edges, either of
armchair or zigzag shape [Fig.~\ref{fig:geom}(b)]. Following recent
estimates for passivation of the dangling carbon bonds at the edges of
graphene samples (e.g., by attached hydrogen),\cite{Louie} we set the
potential of the out-most carbon atoms to 1.7~eV. Although in this
case many of the surface states present in a perfect zizag boundary
remain suppressed, such a confining potential leads to substantial
changes in the energy level spectrum [see Fig.~\ref{fig:EV_hard}(a)]
as compared to the model potential [Eq.~(\ref{eq:pot})]. Most
importantly, in the low-field regime [see Fig.~\ref{fig:EV_hard}(b)]
the linear $B$-field dependence is replaced by a quadratic dependence
($\propto B^2$) of the level splitting resembling the nonrelativistic
diamagnetic response. This is due to the presence of sizable avoided
crossings near $B=0$ as a result of broken valley symmetry caused by the
edges. The valley symmetry is, thus,
broken upon reflection at atomically sharp ``clean'' zigzag edges and $\tau_z$ is
no longer conserved. In terms
of perturbation theory, the confining potential $V$ now includes
off-diagonal components in the pseudospin degree of freedom
\begin{equation}
W = \sigma_x\otimes\expval{H_B}_{KK}\tau_z +
\mathrm{Re}\expval{V}_{KK^{\prime}}\tau_x +
\mathrm{Im}\expval{V}_{KK^{\prime}} \tau_y,\label{E_pert_kk}
\end{equation}
with coupling matrix elements between the valleys
$\expval{V}_{KK^{\prime}} = \sigma_0\otimes \bra\psi_A V \ket\psi_{A^{\prime}}
+ \sigma_x\otimes \bra\psi_A V \ket\psi_{B^{\prime}}$ and
eigenvalues
\begin{equation}
\varepsilon = \pm \sqrt{\expval{H_B}_{KK}^2 + (\Delta\varepsilon/2)^2},
\quad \Delta\varepsilon = 2\left|\expval{V_{KK^{\prime}}}\right|.
\label{EV_pert_kk}
\end{equation}
The coupling between $K$ and $K'$ cones thus leads, according to the
Wigner-von Neumann non-crossing rule,\cite{wignervonneumann} to
avoided crossing with level splittings $\Delta\varepsilon$ [see
  Fig.~\ref{fig:EV_hard}(b)] proportional to the coupling strength
$V_{KK^{\prime}}$ between the two Dirac cones. Conversely, a fit to
Eq.~(\ref{EV_pert_kk}) yields a sensitive indicator for the amount of
\KKp\ scattering in the quantum dot.

For high magnetic fields [see Fig.~\ref{fig:EV_hard}(c)], the presence
of $V_{KK'}$ coupling lead to a large number of correlated avoided
crossings when the edge states move towards the zeroth bulk Landau
level. In other words avoided crossings appear when the energy of
eigenstates evolving towards the $E_0^D$ level ``pass'' through the
energy $E_1^D(B)$ of the first Landau level
[Fig.~\ref{fig:EV_hard}(c)]. Due to a large number of avoided
crossings, there are no states continuously following the first Landau
level. We rather observe a bundle of states sequentially moving along
the characteristic energy of the first Landau level, much like in a
relay [see Fig.~\ref{fig:EV_hard}(c)]. Such an interrelated sequence
of avoided crossings is well-known from atomic physics as ``diabatic
ridge'' - riding states localized on potential
barriers.\cite{Joachim_Ridge} A direct consequence is that the
evolution of eigenstates for an increasing magnetic field [see green
  highlighted line in Fig.~\ref{fig:EV_hard}(a)] features sharp
``kinks'' when crossing the ridge following the first Landau level
[see arrow in Fig.~\ref{fig:EV_hard}(c)]. As the state is transiently
trapped by the ridge, it moves away from the Dirac point, and
continues again monotonically towards the Dirac point once clear of
the ridge. These kinks due to the ridge riding mechanism have been
observed in the experiment\cite{Guttinger09} serving as an additional
indicator for the position of the lowest Landau level and the
electron-hole crossover.

The present results show that the atomically sharp edges destroy
the linear $B$-dependence of the energy levels at weak fields, but
they do preserve the square-root $B$-dependence
at very high fields. The linear $B$ dependence results from the fragile
suppression of \KKp\  scattering ($\propto \expval{V}_{KK'}$) between the Dirac
cones while the square-root dependence results from the much more
robust dispersion relation of the individual cone. Therefore, Landau
levels survive the introduction of sharp boundaries much better than
the energy levels at weak fields. In turn, even when in the experiment
a Dirac-like Landau level spectrum is recorded, many other features of
the same graphene sample may very well show large deviations from predictions
based on Dirac theory.

\subsection{Dots with bulk disorder}
\begin{figure}
\hbox{}\hfill\epsfig{file=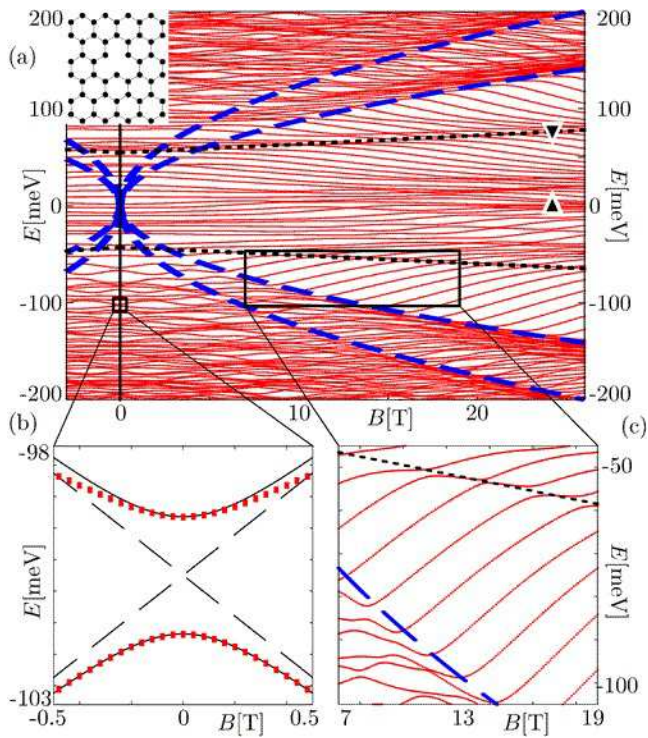, width = 8.5cm}\hfill\hbox{}
\caption{(color online) Same as Fig.~\ref{fig:EV_clean} for a quantum dot featuring
  30 single lattice vacancies (see inset) out of a total number of
  100.000 carbon atoms. Dotted lines mark the evolution
  of two states localized at one defect.  Symbols mark the states for which the
  corresponding wavefunction is shown in Fig.~\ref{fig:ES_disorder}.}
\label{fig:EV_disorder}
\end{figure}
 
\begin{figure}
\hbox{}\hfill\epsfig{file=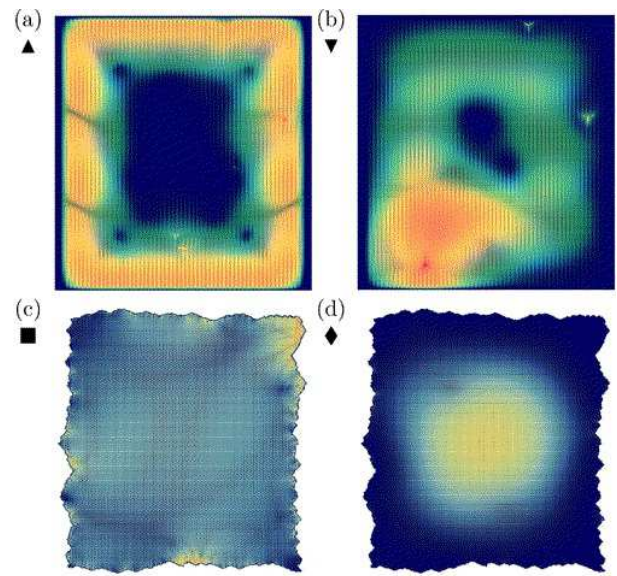, width =
  8cm}\hfill\hbox{}
\caption{(color online) Eigenstates of a graphene quantum dot with
  disorder: ($\blacktriangle$, $\blacktriangledown$) 30 single
  vacancies [one single vacancy shown as inset in
    Fig.~\ref{fig:EV_disorder}(a)] or ($\blacksquare$,
  $\blacklozenge$) edge roughness of $\pm 2$~nm. Positions are marked
  by corresponding symbols in Fig.~\ref{fig:EV_disorder}(a) and
  Fig.~\ref{fig:EV_rough}(a).}
\label{fig:ES_disorder}
\end{figure}

To further elucidate the role of lattice symmetry breaking in graphene
quantum dots, we now consider isolated lattice defects in the
bulk. The preceding results suggest that disorder realizations that
break the \KKp\ symmetry are of crucial importance. We therefore
consider first single lattice vacancies [see inset in
  Fig.~\ref{fig:EV_disorder}(a)] with defect densities of $\pp =
10^{-5}$ to $10^{-3}$ impurities per carbon atom.
%
%
To isolate the effect of such bulk defects from the \KKp\ scattering
of the edges, we first use the smooth symmetry preserving boundary
potential [see Eq.~(\ref{eq:pot})]. Overall, the diamagnetic spectrum
closely resembles that of clean samples with atomically sharp edges
(see Fig.~\ref{fig:EV_disorder}). In particular, avoided crossings
near $B=0$ with a quadratic field dependence as well as the formation
of sequences of avoided crossings along the ridges of bulk Landau
levels are found.  Wave functions of eigenstates at these energies
display patterns which are very similar to those for the clean system
[compare Fig.~\ref{fig:ES_clean}(b) and
  Fig.~\ref{fig:ES_disorder}(a)]. Likewise, the kink pattern at the
crossing of the ridges appears robust against disorder [see,
  e.g.,~Fig.~\ref{fig:EV_disorder}(c)]. We do, however, observe new
ridges between the Landau level energies which were absent for edge
scattering [see dotted lines in Fig.~\ref{fig:EV_disorder}(a,c)].  The
corresponding eigenstates near these new ridges [see
  Fig.~\ref{fig:ES_disorder}(b), marked by $\blacktriangledown$ in
  Fig.~\ref{fig:EV_disorder}(a)] are pinned to a single defect, where
the lattice periodicity and the A-B sublattice symmetry are broken,
and the partitioning of the wavefunction in four components according
to the Hamiltonian in Eq.~(\ref{E_Graph}) fails. Such localized states
can therefore be expected to behave differently from the bulk Landau
levels. The resulting ridges feature a very weak quasi-linear magnetic
field dependence. We therefore conjecture that these structures are
due to avoided crossings with such localized defect states. Recent
analysis\cite{Skrypnyk06, Robinson08} has shown that \KKp\ scattering
at impurities in graphene, i.e.~the strength of $V_{KK'}$,
tends to be strongly energy dependent. Bound
states due to adsorbates (leading to an enhanced local density of
states at the adsorption site) frequently have energies close to the
Dirac point.\cite{Robinson08,Wehling09} Since the localized defect
states with energy $E_n$ feature a very weak explicit linear magnetic field
dependence,
\begin{equation}
  E_n [ V_{KK'}, B] \approx E_n[V_{KK'}] + \alpha \left|B\right|,\quad |\alpha| < 1\frac{\mathrm{meV}}{\mathrm{T}},
\end{equation}
the implicit magnetic field dependence of $V_{KK'}(E_n[B])$ may be
neglected.  While the detailed energy dependence of $V_{KK'}$ may be
connected to the specific defect present (e.g., Stone-Wales defects,
or attached nitrogen molecules), we still qualitatively expect an
analogous linear $B$-field dependence. Such localized states with weak
magnetic field dependence were also observed experimentally in Coulomb
blockade measurements.~\cite{Guttinger09, Guttinger09b}

\begin{figure}
\hbox{}\hfill\epsfig{file=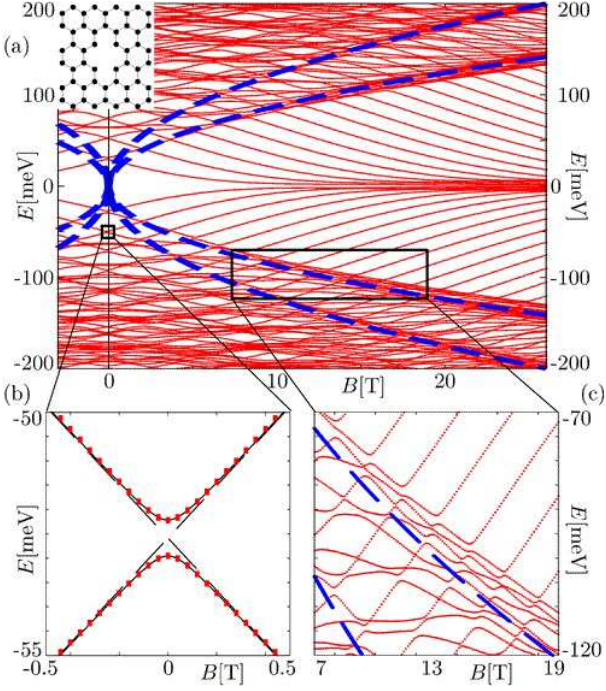, width = 8cm}\hfill\hbox{}
\caption{(color online) Same as Fig.~\ref{fig:EV_clean} for a quantum dot featuring 30
  double lattice vacancies (see inset) in 100.000 carbon atoms.}
\label{fig:EV_impurities}
\end{figure}

In the low field regime, we observe a significant change in the
$B$-evolution of eigenenergies: avoided crossings become asymmetric
[see Fig.~\ref{fig:EV_disorder}(b)]. The reason is that the
\KKp\ splitting introduced by the lattice vacancies is strong enough
to yield different matrix elements for $\expval{H}_{KK}$ and
$\expval{H}_{K^{\prime}K^{\prime}}$: consequently, the slope of both
eigenvalues of Eq.~(\ref{E_pert_kk}) is different [see
  Fig.~\ref{fig:EV_disorder}(b)]. To illustrate that the diamagnetic
spectrum, in particular the avoided crossing distribution, is due to
the breaking of the A-B sublattice and, in turn, to the breaking of
valley pseudospin symmetry induced by the defects, we present as
counter example the spectrum for double vacancies. We introduce such a
vacancy in accordance to our soft-wall potential, Eq.~(\ref{eq:pot}).
As such double vacancies act on an entire unit cell in the hexagonal
lattice, they approximately conserve the electron-hole and
\KKp\ symmetry. We find, indeed, that for the same number of defects
as of single vacancies (Fig.~\ref{fig:EV_disorder}), only avoided
crossings with comparatively small energy splittings appear in the
energy level diagram (Fig.~\ref{fig:EV_impurities}), resembling much
more closely Fig.~\ref{fig:ES_clean}.  This clearly indicates that it
is the breaking of the A-B symmetry and not the presence of defects
per se which is responsible for the break-down of the Dirac picture
for graphene quantum dots.

\subsection{Dots with edge disorder}

\begin{figure}[t]
\hbox{}\hfill\epsfig{file=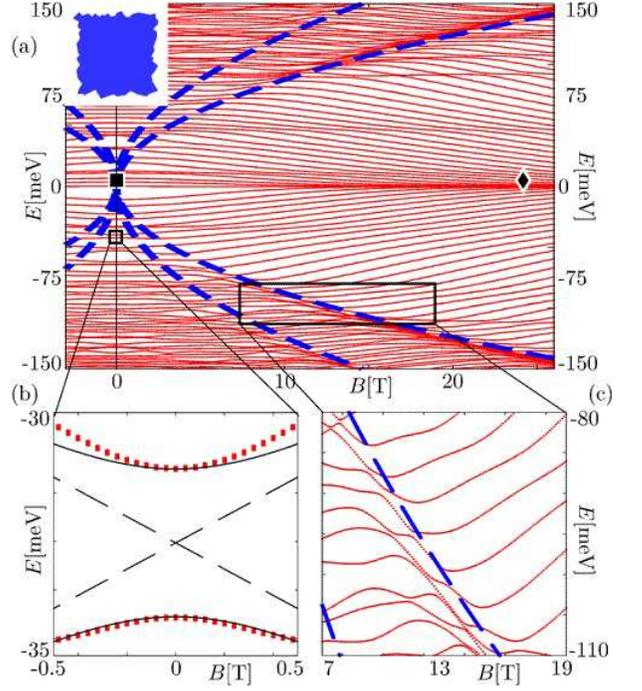, width = 8cm}\hfill\hbox{}
\caption{(color online) Same as Fig.~\ref{fig:EV_hard} for a quantum dot featuring 
  rough edges (with a roughness amplitude $\delta d= \pm 2$~nm).
  Symbols mark the states for which the corresponding wavefunction is shown in
  Fig.~\ref{fig:ES_disorder}.}
\label{fig:EV_rough}
\end{figure}

We consider now a clean graphene quantum dot with atomically sharp but
disordered edges. We connect short straight edge segments (with a
random length between 0.5-3~nm) to obtain a polygon-shaped boundary
[see Fig.~\ref{fig:geom}(c)] with a disorder amplitude of $|\delta d|
\le 2$~nm. $\delta d$ is thus (at energies close to the Dirac point)
smaller than the wavelength $\lambda_{\mathrm F}$ of the confined
particles as well as the magnetic length, but larger than the lattice
constant. Since rough boundaries, just like bulk defects, localize
states,\cite{Wimmer,lib09} we expect similar signatures of these two
types of disorder. Indeed, in the low field regime, dots with rough
edges feature a similar pattern of fluctuating energy levels as dots
with single-lattice defects (compare Fig.~\ref{fig:EV_rough} and
Fig.~\ref{fig:EV_disorder}). The spectra are so similar that it is
difficult to distinguish between bulk and edge disorder breaking the
\KKp\ symmetry. Also in the high-field regime, the evolution towards
the Landau levels features the correlated sequence of avoided
crossings reflecting the diabatic ridges [see
  Fig.~\ref{fig:EV_rough}(c)]. Wave functions of eigenstates at these
energies display patterns which are very similar to those for the
clean system [compare Fig.~\ref{fig:ES_clean}(a) and
  Fig.~\ref{fig:ES_disorder}(d)]. In particular, states near Landau
levels localize in the interior of the dot, and thus are not
influenced by edge disorder. Likewise, the kink pattern at the
crossing of the ridges is robust against edge disorder. While the
inclusion of edge disorder does not give rise to qualitatively new
effects on the eigenenergy spectrum in the high-field regime,
differentiating between edge and bulk disorder might become possible
by probing the different scaling behavior for bulk and edge disorder
with the size of the graphene quantum dot.

\begin{figure}[t]
\hbox{}\hfill\epsfig{file=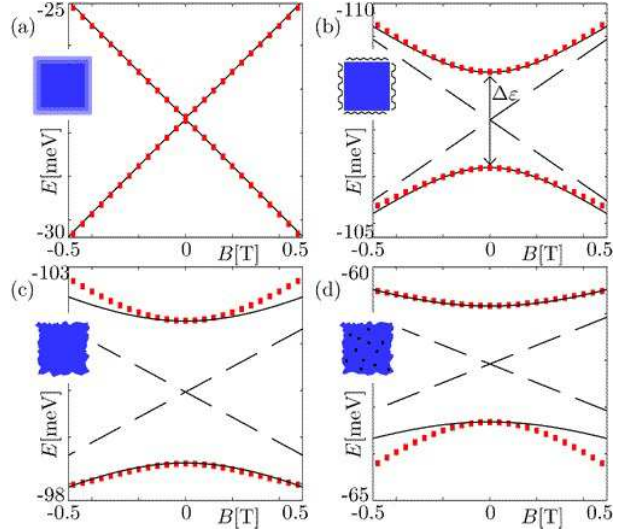,
  width=8cm}\hfill\hbox{}
\caption{(color online) (Avoided) crossings for (a)
  soft edges, (b) hard edges, (c) rough edges, and (d) rough edges plus
  bulk disorder. The level splitting is (a) 0.1~meV, (b) 2~meV,(c) 3~meV,
  and (d) 2.5~meV. }
\label{fig:summary}
\end{figure}

It is instructive to directly contrast the level splitting
$\Delta\varepsilon$ of an avoided crossing [Eq.~(\ref{EV_pert_kk})]
due to finite \KKp\ coupling for different scenarios: (i)
valley-symmetry preserving confinement [Fig.~\ref{fig:summary}(a)],
(ii) clean graphene quantum dots with atomically sharp edges
[Fig.~\ref{fig:summary}(b)] and (iii) disordered graphene quantum dots
[Figs.~\ref{fig:summary}(c,d)]. We observe an increase in the size of
the average level splitting $\expval{\Delta\varepsilon}$ due to
\KKp\ coupling at the edges and impurities. In scenario (i)
$\expval{\Delta\varepsilon}$ is at least one order of magnitude
smaller than the mean level spacing $\expval{\delta\varepsilon}$,
$\expval{\Delta \varepsilon} \ll \expval{\delta \varepsilon}$. In (ii)
where $\expval{\Delta \varepsilon} < \expval{\delta \varepsilon}$ a
first-order perturbative treatment of \KKp\ coupling correctly
describes the level repulsion. For localization at defects (iii) where
$\expval{\Delta \varepsilon} \lesssim \expval{\delta\varepsilon}$
level pairs at $K$ and $K'$ feature different quadratic dependences on
$B$, i.e., they can no longer be parametrized by
Eqs.~(\ref{E_pert_kk}) and (\ref{EV_pert_kk}).

\section{Comparison with experiment: level spacing fluctuations}

\begin{figure}
\hbox{}\hfill\epsfig{file=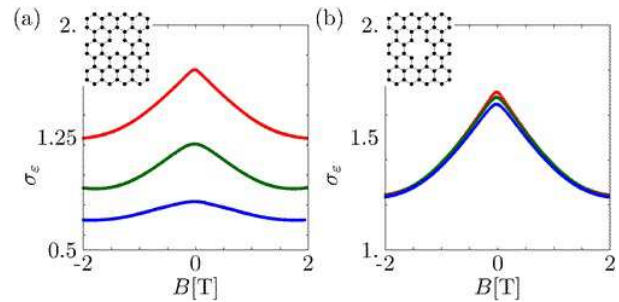, width = 8cm}\hfill\hbox{}
\caption{(color online) Variance $\sigma_\varepsilon$ of the mean level
  spacing [see Eq.~(\ref{eq:sigma})] as a function of magnetic field for (a)
  single-vacancy defects and (b) double-vacancy defects (note the expanded 
  $y$-scale) for three values
  of disorder concentrations: from top to bottom 
$\pp = 3\cdot 10^{-5}, 6\cdot 10^{-5}$ and $2\cdot 10^{-4}$.} 
\label{fig:EV_theor}
\end{figure}

For a comparison with the experimental data for the magnetic response
of graphene quantum dots we pursue two strategies: in a direct
approach we compare our models with the observed parametric $B$-field
evolution of individual Coulomb blockade peaks. Alternatively, we
identify the $B$-field dependence of the level spacing fluctuations
(variance) as a robust measure for the degree of disorder in a
graphene quantum dot. Specifically, we determine the rescaled (or
unfolded) variance $\sigma_\varepsilon$ of the distribution of
neighboring energy level spacings $\delta\varepsilon$ 
\begin{equation}
\sigma_\varepsilon := 
\frac{1}{\expval{\delta\varepsilon}}\sqrt{\expval{(\delta\varepsilon)^2 -
  \expval{\delta\varepsilon}^2}}.
\label{eq:sigma}
\end{equation}
Since switching on a magnetic field $B$ leaves the number of states
unchanged, $\expval{\delta \varepsilon}$ is (approximately)
independent of $B$, while higher moments of the level distribution are
drastically affected. At $B = 0$, pairs of energy levels are split by
the characteristic energy $\Delta\varepsilon$ of the avoided level
crossings. However, as long as the mean width of the avoided crossing
$\expval{\Delta\varepsilon}$ [Eq.~(\ref{EV_pert_kk})] is small
compared to the mean level spacing $\expval{\delta\varepsilon}$, the
level sequence fluctuates between small and large spacings, while
spacings of the order of $\delta\varepsilon$ are unlikely. We thus
expect for magnetic field $B=0$ a comparatively large variance
$\sigma_\varepsilon$. For increasing $|B|$ the levels become more
equally spaced, leading to a decrease in $\sigma_\varepsilon$.
Correspondingly, the variance $\sigma_\varepsilon$ of the level
spacings should feature a peak at $B=0$. The numerical results for the
dependence of $\sigma_\varepsilon$ on $B$ for different disorder
strength (i.e., different $\left|V_{KK'}\right|$) are shown in
Fig.~\ref{fig:EV_theor}. Our data display a pronounced peak of
$\sigma_\varepsilon$ for the clean flake slowly decreasing for
increasing number of single-vacancy defects (i.e., for increasing
\KKp\ scattering) [see Fig.~\ref{fig:EV_theor}(a)].  Note that both
the peak height at $B=0$ and the overall value of $\sigma_\varepsilon$
decrease with increasing disorder.  The latter can be explained by the
emerging localized states that feature a regular spacing (and hence a
suppressed variance $\sigma_\varepsilon$). Consequently, if a given
spectral region is more prone to feature localized states due to
adsorbates than others,\cite{Robinson08, Wehling09} we expect an
energy-dependence on $\sigma$ depending on the specific type of
adsorbate. Such an energy dependence could be exploited to directly
measure the energy dependence of \KKp scattering by determining the
statistics of Coulomb blockade resonances at different back-gate
voltages. For comparison we also show $\sigma_\varepsilon$ for double
vacancies preserving A-B symmetry. [see Fig.~\ref{fig:EV_theor}(b),
  note the different scales]. Accordingly, $\sigma_\varepsilon$ is,
indeed, strongly dependent on the amount of \KKp\ scatterers, not on
the overall number of defects.

\begin{figure}
\hbox{}\hfill\epsfig{file=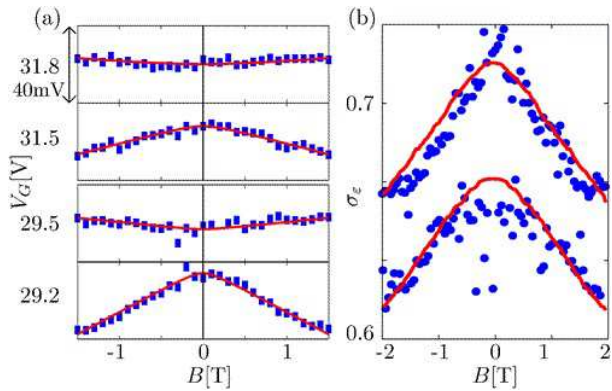,
  width=8cm}\hfill\hbox{}
\caption{(color online) (a) Coulomb blockade peaks measured as a
  function of applied plunger gate electrode potential ($V_G$) and
  magnetic field applied perpendicular to the sample
  (see Ref.~[\onlinecite{endnote1}] for details about the
  measurement). Lines: fit to Eq.~(\ref{EV_pert_kk}) for two measured
  Coulomb blockade peak pairs with a level splitting at $B=0$ of
  $\approx 0.3 V$. (b) Normalized variance of the level spacing
  $\sigma_\varepsilon$ [see Eq.~(\ref{eq:sigma})] as a function of
  magnetic field. Dots denote an average over 20 consecutive
  experimental Coulomb blockade peaks for two values of back gate
  voltage $V_{BG}$ [corresponding to the upper ($V_{BG} =$ 20 V) and lower
    ($V_{BG} =$ 38 V) data points]. Solid lines represent simulations for
  different edge roughness amplitude $\delta d$ 
    [Fig.~\ref{fig:geom}(d)]
  as only fit parameter: $\delta d\approx $0.5 (0.6) nm, for the upper
  (lower) curve.}
\label{fig:experiment}
\end{figure}

The decrease in peak height with increasing \KKp\ scattering should thus
provide a robust and sensitive measure for \KKp\ scattering present in the
experiment.
To test this conjecture, we have measured the evolution of 42 Coulomb
blockade peaks for varying magnetic field.\cite{endnote1} We follow the
parametric motion of the peak positions, and hence the eigenstate
energies, as a function of magnetic field $B \in [-2,2]$T. To compare
with our numerical results, we take into account a charging energy of
13~meV (determined independently\cite{endnote1}) as well as spin (by
Zeeman splitting). We observe, indeed, a quadratic $B$-field
dependence rather than the linear dependence predicted for conserved
\KKp\ symmetry of the Dirac equation. Our experimental data can be
well described by Eq.~(\ref{EV_pert_kk}) [see
  Fig.~\ref{fig:experiment}(a)].  For pairs of consecutive Coulomb
blockade resonances belonging to the same avoided crossing, we find a
mismatch in slopes, in agreement with our numerical findings for the
rough-edged quantum dot [compare Fig.~\ref{fig:experiment}(a) and
  Fig.~\ref{fig:summary}(c,d)]. This has important consequences for
the interpretation of experimental data: the roughness present in the
experimental dot does not allow to disentangle $K$ and $K'$ states. To
be more quantitative, we compare our simulations for the level
variance with experimental data. We indeed find good agreement as
confirmed by a noticeable peak also in the experimental data for
$\sigma_\varepsilon(B)$ [see Fig.~\ref{fig:experiment}(b)]. The offset
between the two data sets in Fig.~\ref{fig:experiment}(b) is
attributed to a possible energy dependence of $K-K'$ scattering
as well as statistical fluctuations in charging energy and in the number
of localized states for different values of the back gate voltage. By
using the edge roughness $\delta d$ as the only adjustable parameter,
we can match the measured $\sigma_\varepsilon(B)$ very well with our
numerical simulation [see lines in Fig.~\ref{fig:experiment}(b)]. Good
agreement is found for an edge roughness of about $\delta d \approx
0.5\pm0.2$~nm, or, equivalently, a \KKp\ scatterer concentration in the bulk $n_i
\approx (3.5\pm 1 )\cdot 10^{-4}$, both of which are well within
expectation.  We emphasize that, although we can quantify the
resulting overall strength of \KKp\ coupling in our experimental
quantum dot, we cannot disentangle whether the observed
\KKp\ scattering comes from edge roughness, lattice defects, or
disorder through flake-substrate interactions with a length scale
comparable to the lattice constant.

\section{Conclusions}

We have investigated the evolution of eigenstates in graphene quantum
dots with increasing magnetic field. Concentrating on the energy
regime around the Dirac point, we observe a smooth transition from a
linear density of states to the emergence of Landau levels.  At high
field strength, we find that Landau levels follow the square-root
dependence of the Dirac equation, manifested in the energy level
diagram by sequences of correlated avoided crossings along ``diabatic
ridges''. These ridges lead to characteristic kinks in the evolution
of energy states that cross a Landau level. Appearing also in Coulomb
blockade measurements, these kinks can be used to experimentally pin
down the electron-hole crossover point.\cite{Guttinger09} In the
perturbative regime of small magnetic fields, we find that the linear
dependence on $B$ predicted by the model of massless Dirac fermions
disappears when the valley symmetry is broken. Even perfect armchair
and zigzag edges are sufficient to break the sublattice symmetry
giving rise to avoided crossings with a quadratic dependence on $B$
instead. A similar effect is observed for lattice defects: single
lattice vacancies break the \KKp\ symmetry and thus result in
avoided crossings with substantial level splittings (even for defect
concentrations as low as 1 in 20.000).  By comparison with double
vacancies which conserve the sublattice symmetry we show 
that it is not the presence of disorder per se which leads to
deviations from predictions by the Dirac equation, but the breaking of
valley symmetry.

We compare our theoretical predictions with experimental results on
the parametric $B$-field evolution of Coulomb blockade peaks. As a
quantitative indicator for the strength of \KKp\ scattering, we
identify the variance $\sigma_\varepsilon$ of the level spacing
distribution. We observe a peak in the variance at $B=0$ due to level
correlations near avoided crossings. We find quantitative agreement
between the measured and the calculated data for $\sigma_\varepsilon$
which enables us to pinpoint the amount of \KKp\ scattering present in
our experimental flake. The present results provide a sensitive
indicator for the quality of the graphene dot and demonstrate the
limits of the Dirac picture in describing the experiment.

\begin{acknowledgments}
We thank K.~Ensslin, T.~Ihn, and M. Wimmer for valuable discussions.  F.L. and
J.B. acknowledge support from the FWF-SFB 016 and FWF-SFB 041,
J.G. from the Swiss National Science Foundation, C.S. support from
NCCR. Computational resources from the Vienna Scientific Cluster (VSC)
are gratefully acknowledged.
\end{acknowledgments}


\begin{thebibliography}{99}



\bibitem{pon08} L. A. Ponomarenko, F. Schedin, M. I. Katsnelson,
  R.~Yang, E.~H.~Hill, K.~S.~Novoselov, A.~K.~Geim, Science {\bf 320},
  356 (2008)

\bibitem{sta08a} C. Stampfer, J. G\"uttinger, F. Molitor, D. Graf,
  T. Ihn, and K. Ensslin, Appl. Phys. Lett.~{\bf 92}, 012102 (2008),
  C. Stampfer, E. Schurtenberger, F. Molitor, J. G\"uttinger, T. Ihn,
  and K. Ensslin, Nano Lett.~{\bf 8}, 2378 (2008)

\bibitem{sch09} S. Schnez, F. Molitor, C. Stampfer, J. G\"uttinger,
  I. Shorubalko, T. Ihn, and K. Ensslin, Appl. Phys. Lett.~{\bf 94},
  012107 (2009)

\bibitem{gue08} J. G\"uttinger, C. Stampfer, S. Hellm\"uller,
  F. Molitor, T. Ihn, and K. Ensslin, Appl. Phys. Lett.~{\bf 93},
  212102 (2008)

\bibitem{han07}
M. Y. Han, B. \"Ozyilmaz, Y. Zhang, and P. Kim, Phys. Rev. Lett.~{\bf 98}, 206805 (2007)

\bibitem{che07}
Z. Chen, Y.-M. Lin, M. Rooks and P. Avouris, Physica E,~{\bf 40}, 228 (2007)

\bibitem{dai08}
X. Li, X. Wang, L. Zhang, S. Lee, H. Dai, Science {\bf 319}, 1229 (2008)

\bibitem{tod09}
K. Todd {\it et al.}, Nano Lett. {\bf 9}, 416 (2009)

\bibitem{liu09}
X. Liu, J. B. Oostinga, A. F. Morpurgo, and L. M. K. Vandersypen, Phys.~Rev.~B~{\bf 80}, 121407(R) (2009)

\bibitem{mos09}
J. Moser, and A. Bachtold, Appl. Phys. Lett. \textbf{95}, 173506 (2009)
\bibitem{sta09} C. Stampfer, J. G\"uttinger, S. Hellm\"uller,
  F. Molitor, K. Ensslin, and T. Ihn, Phys. Rev. Lett.~{\bf 102},
  056403 (2009)

\bibitem{rit09}
K. A. Ritter and J. W. Lyding, Nat. Mater.~{\bf 8}, 235 (2009)

\bibitem{gei07}
A.~K. Geim and K.~S. Novoselov, Nat. Mater.~{\bf 6}, 183 (2007)

\bibitem{kat07}
M.~I. Katnelson, Materials Today~{\bf 10(1-2)}, 20 (2007)

\bibitem{tra07}
B. Trauzettel, D.~V. Bulaev, D.~Loss, and G.~Burkard, Nature Physics~{\bf 3}, 192 (2007)

\bibitem{kan05} 
C. L. Kane and E. J. Mele, Phys. Rev. Lett.~{\bf 95}, 226801 (2005)

\bibitem{min06} H. Min {\it et al.}, Phys. Rev. B~{\bf 74}, 165310 (2006)

\bibitem{hue06} 
D. Huertas-Hernando, F. Guinea, and A. Brataas, Phys. Rev. B~{\bf 74}, 155426
(2006)


\bibitem{ber87}
M. V. Berry and R. J. Mondragon, Proceedings of the Royal Society of London, A~{\bf 412},
53-74 (1987)

\bibitem{sch08}
S. Schnez, K. Ensslin, M. Sigrist, and T. Ihn, Phys. Rev. B,~{\bf 78} 195427 (2008)

\bibitem{rec07}
P. Recher, B. Trauzettel, A. Rycerz, Ya. M. Blanter, C. W. J. Beenakker and A. F. Morpurgo,
Phys. Rev. B,~{\bf 76}, 235404 (2007)

\bibitem{rec09}
P. Recher, J. Nilsson, G. Burkard, and B. Trauzettel, 
Phys. Rev. B,~{\bf 79}, 085407 (2009)
\bibitem{Wimmer}
M. Wimmer, A. R. Akhmerov, and F. Guinea, arXv:1003.4602

\bibitem{lib09}
F. Libisch, C. Stampfer, and J. Burgd\"orfer, Phys. Rev. B~{\bf 79}, 115423 (2009)

\bibitem{you09}
A. F. Young and P. Kim, Nature Phys.~{\bf 5}, 222-226 (2009)

\bibitem{cas08} For review see, e.g., A.~H.~Castro Neto, F.~Guinea,
  N.~M.~Peres, and A.~K.~Geim, Rev. Mod. Phys.~{\bf 81}, 109 (2009)


\bibitem{Haldane88} F. D. M. Haldane, Phys. Rev. Lett. {\bf 61}, 2015 (1988)
\bibitem{Zheng02} Y. Zheng and T. Ando, Phys. Rev. B {\bf 65}, 245420 (2002)
\bibitem{McCann06} E. McCann and V. I. Fal'ko, Phys. Rev. Lett. 96, 086805 (2006)
\bibitem{Sheng07} L. Sheng, D. N. Sheng, F. D. M. Haldane, and L. Balents, 
Phys. Rev. Lett. {\bf 99} 196802 (2007)
\bibitem{Ostrovsky08} P. M. Ostrovsky, I. V. Gornyi, and A. D. Mirlin, 
Phys. Rev. B {\bf 77}, 195430
\bibitem{Schulz09a} 
D. A. Bahamon, A. L. C. Pereira, and P. A. Schulz, Phys. Rev. B {\bf 79}, 125414 (2009)
\bibitem{Schulz09b}
P. H. Rivera, A. L. C. Pereira, and P. A. Schulz, Phys. Rev. B {\bf 79}, 205406 (2009)


\bibitem{nov05} K. S. Novoselov, A. K. Geim, S. V. Morozov, D.~Jiang,
  M.~I.~Katsnelson, I.~V.~Grigorieva, S.~V.~Dubonos, A. A. Firsov,
  Nature~{\bf 438}, 197-200, (2005).

\bibitem{zha05}
Y. Zhang, Y.-W.~Tan, H.~L.~Stormer, P.~Kim, Nature~{\bf 438}, 201-204, (2005).

\bibitem{dom99}
N. Dombay, and A. Calogeracos, Phys. Rep.~{\bf 315}, 4158 (1999)

\bibitem{kat06}
M. I. Katsnelson, K. S. Novoselov, amd A. K. Geim, Nature Phys.~{\bf 2},
620625 (2006)

\bibitem{Semenoff}
G. W. Semenoff, Phys. Rev. Lett. {\bf 53}, 2449 (1984)

\bibitem{Guttinger09} J. G\"uttinger, C. Stampfer, F. Libisch,
  T. Frey, J. Burgd\"orfer, T. Ihn and K. Ensslin,
  Phys. Rev. Lett. {\bf 103}, 046810 (2009).

\bibitem{Sasaki06} K. Sasaki, S. Murakami, and R. Saito, 
App. Phys. Lett. 88, 113110 (2006)

\bibitem{McCann06b} E. McCann, K. Kechedzhi, V. I. Fal'ko, H. Suzuura, T. Ando, and
B. L. Altshuler, Phys. Rev. Lett {\bf 97}, 146805 (2006)

\bibitem{Rabi}
I. I. Rabi, Zeitschrift f. Physik A {\bf 49}, 507 (1928)

\bibitem{McClure56} J. W. McClure, Phys. Rev. {\bf 104}, 666 (1956)

\bibitem{Wallace47} P. R. Wallace, Phys. Rev. {\bf 71}, 622 (1947)


\bibitem{TBReich}
S.~Reich, J. Maultzsch, C. Thomsen, and P. Ordej\'on, Phys.~Rev.~B {\bf 66}, 035412 (2002)
\bibitem{TBWirtz}
A.~Gr\"uneis { \it et al.~}, Phys.~Rev.~B {\bf 78}, 205425 (2008)

\bibitem{Slonc}
J. C. Slonczewski and P. R. Weiss, Phys. Rev. {\bf 109}, 272 (1958)

\bibitem{adam02}S. Adam, M. L. Polianski, X. Waintal, and
  P. W. Brouwer, Phys. Rev. B {\bf 66}, 195412 (2002).

\bibitem{lent91} C. S. Lent, Phys.~Rev.~B {\bf 43}, 4179 (1991).
\bibitem{Lanczos}
C. Lanczos, Journal of Research of the National Bureau of Standards {\bf 45},
255, (1950)
\bibitem{MUMPS}
P. R. Amestoy, I. S. Duff, J. Koster, and J.-Y. L'Excellent,
SIAM Journal of Matrix Analysis and Applications
{\bf 23}, 15-41 (2001).
P. R. Amestoy, A. Guermouche, J.-Y. L'Excellent and
S. Pralet, Parallel Computing {\bf 32}, 136-156 (2006).

\bibitem{wignervonneumann} J. von Neumann and E. P. Wigner, Z. Physik
  {\bf 30}, 467, 1929.

\bibitem{Louie}
Y.-W. Son, M. L. Cohen, and S. G. Louie, Phys. Rev. Lett. {\bf 97}, 216803, (2006).

\bibitem{Joachim_Ridge} P. S. Krsti\'c, C. O. Reinhold, and
  J. Burgd\"orfer, Phys. Rev. A {\bf 63}, 052702 (2001). See also
  J. Burgd\"orfer, N. Rohringer, P. S. Krsti\'c, and C. O. Reinhold,
  Nonadiabatic processes near barriers in 'Nonadiabatic transition in
  quantum systems' edited by V.I. Osherov, L.I. Ponomarev,
  Chernogolovka, Russia p. 205 (2004). Institute of Chemical Physics
  RAS ISBN 5-901675-48-7


\bibitem{Skrypnyk06} Yu. V. Skrypnyk and V. M. Loktev, Prys. Rev. B {\bf 73}, 241402(R) (2006)
\bibitem{Robinson08} J. P. Robinson, H. Schomerus, L. Oroszl\'any, and V. I. Fal'ko,
Phys. Rev. Lett. {\bf 101}, 196803 (2008)
\bibitem{Wehling09} T. O. Wehling, M. I. Katsnelson, and A. I. Lichtenstein,
Phys. Rev. B {\bf 80}, 085428 (2009)

\bibitem{Guttinger09b}
J. G\"uttinger, C. Stampfer, T. Frey, T. Ihn, and K. Ensslin,
Physica Status Solidi B {\bf 246}, 2553 (2009) 

\bibitem{endnote1} The measured
  graphene quantum dot had a similar geometry as the dot described in
  Refs.~\onlinecite{Guttinger09, Guttinger09b}. By changing the potential of the back gate electrode
  ($V_G$) a region of suppressed current between $V_G = 10$~V and 45~V was
  found. From measurements of Coulomb blockade peaks in this region (with
  $\approx$ 150 Coulomb peaks) a charging energy of $\approx$13~meV was
  extracted. For more details about the measurements see
  Refs.~\onlinecite{Guttinger09,Guttinger09b}.
\end{thebibliography}
\end{document}